\documentclass[twocolumn,showpacs,preprintnumbers,amsmath,amssymb,pre]{revtex4}

\usepackage{graphicx}
\usepackage{dcolumn}

\begin{document}


\title{A new statistical mechanical formalism for gases}

\author{Ren\'e D. Rohrmann}
 \altaffiliation[Present address: ] { Observatorio Astron\'omico, UNC, 
Laprida 854, (5000) C\'ordoba, Argentina}
\affiliation{Instituto de Astronom\'{\i}a, UNAM, 
A.P. 70-264, 04510 M\'exico D.F., M\'exico}

\date{\today}

\begin{abstract}
An equilibrium theory of classical fluids based on the space distribution 
among the particles is derived in the framework of the energy minimization 
method. This study is motivated by current difficulties
of evaluation of optical properties in atmospheres of degenerate stars.
Present paper focuses on diluted one-component systems, where the interaction 
energy is calculated as a sum of binary contributions. 
The spatial configuration of the gas is described in terms of a particle-state 
variable $v$ which roughly measures the free space surrounding each particle. 
The formalism offers a unified treatment of both thermodynamics and
structure of fluids, since it not only provides the state function (the 
Helmholtz free energy) of a fluid, but also automatically gives information 
on the microstructure of the system (e.g. the nearest-neighbor distribution 
function).
Equations of state and nearest-neighbor distribution functions of perfect 
and hard-body fluids are obtained in straightforward way in one, two and 
three dimensions.
The formalism is applied to describe the atomic population in a partially 
ionized hydrogen gas. Combined and self-consistent evaluations of atomic 
populations and internal effects on bound states are performed in a 
detailed form. 
The present theory allow us to resolve the atomic density at each internal 
level into groups of atoms which experiment different perturbation 
intensities according to the size of their spaces $v$. 

\end{abstract}


\maketitle

\section{ Introduction} \label{s.intr}

The evaluation of the thermodynamical, transport and optical properties of
non-ideal plasmas is of great interest to characterize the behavior of the
matter at different physical conditions. It has many applications in material 
sciences, geophysics and astrophysics. 
In particular, the equations of state provide the basic
thermodynamic quantities needed to determine the physical properties of 
stellar envelopes and interiors. Models of gases must also provide
detailed atomic and molecular populations required to obtain the monochromatic 
opacity, a quantity which is essential to calculate realistic spectra emitted 
from astrophysical objects. Our study is motivated by current difficulties
of evaluation of gas opacities in envelopes of white dwarf stars (WDs).

While non-ideal effects on equations of state (e.g., departures from the 
ideal $P$-$T$-$\rho$ relationship) become present at densities near to 
ionization pressure (which occurs above $\rho \approx 0.3$ g cm$^{-3}$ for 
a pure hydrogen plasma \cite{Fonta}), signatures of non-ideal effects (i.e., 
particle interactions) are evident in spectra emitted by gases at very low 
densities.
As it is well-known since the earliest days of the quantum theory, 
interparticle perturbations are responsable for the reduction of the ionization
energies, shifts of lines and modifications of their profiles \cite{Marge}. 

The reduction of the ionization potential along with the line merging, yields 
the advance of the series limits in the spectrum and forms `pseudocontinuums', 
which resemble extensions of the bound-free adsorptions toward longer wavelength
\cite{Grie64}. Pseudocontinuum opacities may produce strong modifications 
in the spectra of cool WDs with rich-hydrogen envelopes 
\cite{Ber91, Ber97}. However, so far, 
the model atmospheres fail to reproduce in detail the photometric 
observations of WDs with effective temperatures below $4000$ K, and the source 
of discrepancies likely lies in the computation of gas opacities which are
affected by non-ideal effects \cite{BerLeg}.

Current optical simulations \cite{DAM, HHL}, commonly used in the evaluation of 
WD spectra (e.g. \cite{Ber91, Ber97, Lanz, RSAB}), 
are based on the occupation probability formalism elaborated by
Hummer and Mihalas (HM) \cite{HM}. On the basis of earlier works
\cite{ealiest_w} and the implementation of the technique of free 
energy minimization, the HM formalism assigns to each atomic state
$j$ a probability $w_j$ of finding the atom in this state relative to that of 
finding it in a similar ensemble of non-iteracting particles. 
For the computation of optical properties, the quantity $w_j$ is considered 
as an estimate of the number of states of type $j$ that are available to be 
occupied ({\em bound level}), whereas $1-w_j$ is a measure of the fraction of 
$j$-states that are severely perturbed by plasma interactions, such that an 
atom in this fraction is actually unbounded ({\em dissolved level}) \cite{DAM}.
In this heuristic scheme, a radiative excitation of an atom from a bound 
level to a dissolved level constitutes a pseudo-continuum absorption. 
Althought this proposal seems intuitively right, we have recently showed that
it is not self-consistent with the HM formalism \cite{RSAB}. 
Briefly, the optical simulations assume the existence of a fraction of atoms 
with perturbed electronic energies (values shifted relative to the continuum), 
whereas the occupation HM formalism has been developed for particles species 
which have unperturbed energies (ionization potential of isolated atom). 
As a consequence, the atomic population which contributes to a
pseudocontinuum absorption is overvaluated, because the abundances of 
perturbed atoms are calculated with lower bound energies than those assumed 
in their transitions to dissolved states.
This may explain why such optical simulations yield a huge (unphysical) 
pseudocontinuum-Lyman opacity for cool WDs \cite{Ber97}. Since the starting 
level of these processes is the fundamental state, the overevaluation of the 
Lyman pseudocontinuum is more severe than those of Balmer and Paschen.

While a group of gas properties (equations of state and response functions, 
e.g., specific heats) may be calculated directly from the global distribution 
of particles (mean densities), others, such as emissivities and opacities in 
general depend on particular population subsets.
As an appropriate example, we mention the point of view adopted by the 
statistical theory of line broadening by pressure effects. There, the
spectrum of a broadened line is determinated from a statistical sum, where
the contribution of each radiating atom depends on the configuration of its
surrounding particles \cite{margenau}.
This picture can be also applied to continuum transitions.
Thus, an appropriate treatment of pseudo-continnum opacities could be 
developed on the basis of a formalism able to distinguish groups 
of atoms which, at a given instant, experiment different degrees of 
interparticle perturbation. 

Motivated by these considerations, we believe it is physically significant
and useful to develop a statistical mechanical formalism for gases based on 
the distribution of the physical space among the particles. The central 
idea is that to each particle is assigned a space region $v$ nearly 
related with the closeness of particle neighbors, and which is treated as a
configurational parameter in the thermostatistical description of the gas.
A particle with small space $v$ will have very near neighbors and, 
therefore, it will be subjected on the average to high perturbations, while
a particle will be roughly isolated if it has a huge space $v$ assigned. 
This description offers an appropriate framework for the study of
optical properties of gases.
So far, the application of statistical mechanics in the analysis of space
partition into a set of volumes has been circumscribed to some particular 
systems such as cristals \cite{rivier}. Similar theory for gases has not
been formulated as far as we are aware.
As a first step, the formalism developed in this paper is devoted to 
one-component gases at the limit of low densities.

We formulate the theory in terms of the widely used {\em free-energy 
minimization method} \cite{graboske}. 
Our aim is therefore to develop a Helmholtz free-energy model for gases which 
contains a notion of space per particle, and to find
the statistical equilibrium state and the thermodynamic properties of the 
system from the energy minimization with respect to the particle number 
density, subjected to specific conditions. Since the present work concerns
gases at low densities, we assume factorization of the free energy into 
configurational, translational and particle-interaction contributions.

The paper is organized as follows.
Sec. \ref{s.sp_st} briefly reviews the importance of spatial statistical
in numerous research areas, and focuses on the application of
space partition schemes to the analysis of spatial micro-structures in gases.
In Sec. \ref{s.ther} we introduce the volume distribution among particles as
a new configurational parameter in thermostatistics of gases.
The gas is treated with the Maxwell-Boltzmann statistics and the particle 
interactions are represented by additive pair potentials.
The equilibrium states are derived in Sec. \ref{s.equ_sta}. 
In Secs. \ref{s.ideal}-\ref{s.hard} the theory is applied to perfect and 
hard-sphere gases. We show that they have the expected thermodynamic 
properties. There, the Van der Waals model is introduced 
as an illustration of some of the considerations underlying this article.
In Sec. \ref{s.atoms} we explore the application of the theory to the
description of atoms perturbed by plasma effects in a partially ionized,
hydrogen gas. Concluding remarks are given in Sec. \ref{s.conclus}.

\section{ Spatial Statistics} \label{s.sp_st}

One of the most basic properties of a collection of objects is the division 
of space among them. The partition of space is commonly associated to 
processes that create random geometries. There is a considerable variety 
of physical systems which show stochastic structures. For example:
biological tissues \cite{biol}; 
landscape geometries in agriculture and forestry \cite{forestry};
territoriality zoology \cite{hase}; grains in 
foams and metallurgial aggregates \cite{rivier, mulheran}; intermolecular 
cavities in fluids \cite{sastry}; solar granulation \cite{carolus}; 
fragmentation of interestellar clouds \cite{kiang}; stellar populations 
\cite{pasztor}; matter and void distributions in the large-scale structure 
of the Universe \cite{icke}.
Most of these systems have often been studied with computer simulation of 
stochastic, space-filling structures. A variety of geometric constructions 
have been applied in these analyses \cite{foot1}.

Gases in thermodynamic equilibrium are also typical disordered systems.
In fact, althougth macroscopically homogeneous in the absence of external 
fields, these systems have spatial inhomogeneities at microscopic levels. 
The particles of a fluid move continuously and the instantaneous spatial 
concentration of material is non-uniform and fluctuating at different scales.
\begin{figure}
\scalebox{0.43}{\includegraphics{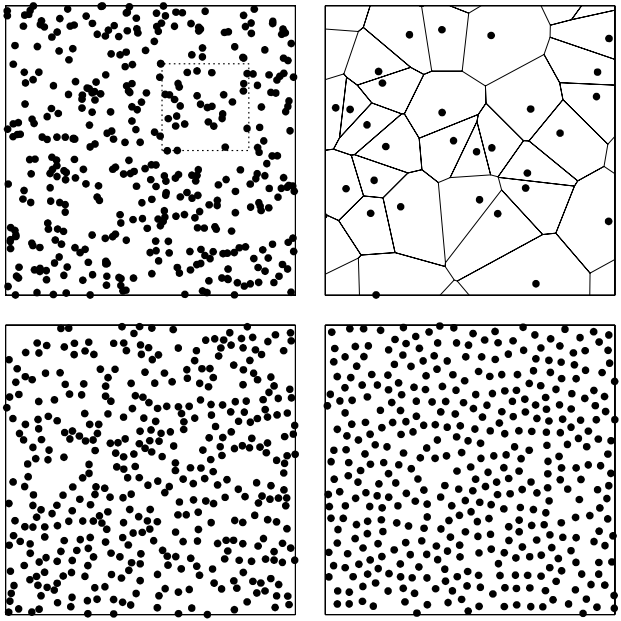}}
\caption{\label{f.f2D} {\em Above}: Poisson distribution of 400 points in a
plane ({\em on the left}) and the Vorono\"{\i} tessellation 
corresponding to points within the dotted box 
({\em on the right}). Solid lines denote the boundaries of the region of 
space closest to each particle.
{\em Below}: Typical configuration of centers of 400 hard discs for a 
packing fraction (fraction of surface covered by the discs) $\eta=0.2$ 
({\em on the left}) and $\eta=0.5$ ({\em on the right}).}
\end{figure}
Fig. \ref{f.f2D} (above on the left) shows a group of objects random 
distributed in a surface, 
which represents a typical configuration of a two-dimensional ideal gas at 
a certain instant. We note that some particles are close enough among them 
while others are separated from their nearest neighbors by larger distances.
One can then imagine a division of the space into a set of regions, one region 
per particle taking into account the closeness of its neighbors.
The most likely well-known partition of space is given by the Vorono\"{\i} 
tesselation \cite{voronoi}. 
The {\em Vorono\"{\i} region} of a given object is the collection of 
points in space nearer to that object than to any other. The complete set of 
Vorono\"{\i} regions form the so-called Vorono\"{\i} diagram of the system. 
An example is given in Fig. \ref{f.f2D} (above on the right).
There are, however, many other types of space divisions (e.g., 
generalizations of the Vorono\"{\i} tesselations \cite{okabe}).

It should be also observed that while the particle distribution of a perfect 
gas corresponds to a Poisson distribution of points in the space, gases formed 
by interacting particles are characterized by other different distributions.
Thus, for example, the well-known one component plasma (model of 
discrete positive charges in a continuous negatively charged background
\cite{baus}) shows a long-range ordered (or `superhomogenous') arrangement 
of particles, which can be appreciated from the analysis of the so-called 
power spectrum of the particle distribution \cite{gabrielli}. 
Also the hard-body fluids show particular spatial patterns which have been
studied extensively \cite{hard_studies}.
As an example, the lower panels in Fig. \ref{f.f2D} show typical configurations 
of hard-disc fluids.
It is evident that the density fluctuations of a gas in 
equilibrium depend on the type of interactions among the particles. 
Furthermore, it is expected that the microscopical spatial structure of a 
gas in equilibrium corresponds to the most probable one according to the 
principles of the statistical mechanics.

The present work is based on the assumption that the microscopical structure 
of a gas can be described by some class of partition of space among the 
particles, and that the most probable partition can be inferred using the 
formalism of the classical statistical mechanics.
The region assigned to a particle, denoted $v$, will be called the 
{\em available volume} of the particle because it gives a notion of 
the size of the particle-free space which surrounds it \cite{v_def}. 
We have mentioned that the division of the space occupied by the gas is not 
unique and, therefore, an exact definition of $v$ is non-trivial in advance.
However, it is possible to establish a statistical formalism of the space
partition in a gas without an explicit description of the volume $v$.
The application of the theory to the ideal gas allows us to identify the 
available volume in this simple system. 
This solution will provide a reference for understanding more complex systems 
composed by interacting particles.

\section{ Gas Description} \label{s.ther}

We consider a gas composed of $N$ particles in a $D$-dimensional container of
size $V$. The region $V$ is a segment, a surface or a volume for 
$D=1$, 2 or 3, respectively, although it will be called generically a volume. 
At low densities the state of the gas can be described using one-particle 
states.
Since we are specially interested in the configurational state of the gas, 
the translational description is here omitted and its well-known contribution 
to the fluid free-energy will be added later.

The physical space occupied by the gas at a given instant is distributed 
among the $N$ particles. To each particle a certain volume $v$ is assigned 
(so-called the available volume, see \S \ref{s.sp_st}) of the total space
$V$. The statistical study of space partition can be performed based only on
the constrains of normalization (there is a volume $v$ per particle) 
and space-filling (the sum of all of them equals the total gas volume),
without an exact specification of $v$ in advance.
We represent the configurational state of the gas by an occupation 
number distribution $N_v$, where $N_v dv$ is the number of particles 
with available volumes between $v$ and $v+dv$.
The physically allowed distributions must satisfy the following conditions
\begin{equation} \label{e.Ncons}
N = \int_0^V N_v \,dv  \;,
\end{equation}
\begin{equation} \label{e.Vcons}
V = \int_0^V v\, N_v \,dv \,.
\end{equation}

{\em Configurational entropy.}
The configurational entropy of the gas is represented by the Boltzmann-Planck 
expression $S_{conf}=k\ln W$, where $k$ is Boltzmann's constant and $W$ 
the number of configurational microstates of the gas which reproduce a 
specific state $N_v$.
Following the usual procedure, we subdivide the total volume into small cells 
which contain fractions $dw$ of neighboring states. In order to evaluate $dw$ 
we turn the attention to well-known results of statistical mechanics. 
There are $d{\bf l}=V d^D p/h^D$ possible 
states for a particle confined in the volume $V$ having a momentum 
${\bf p}$ within a momentum volume $d^D p$ ($h$ is Planck's constant). 
The number of states that corresponds to a physical volume $d^D x$ is 
\begin{equation} \label{e.dw}
state\; number \; in \; (d^D x,d^D p) = \frac {d^D x} 
V d{\bf l} \,.
\end{equation}
This result suggests that the fraction of configurational states of a particle 
having an available volume between $v$ and $v+dv$ (and any momentum) is 
proportional to $dv$ with a proportionality factor $V^{-1}$. 

Thus, a cell has a fraction of states $dv/V$ and contains $N_v \, dv$ 
particles for a given occupation set $N_v$. It is readily shown by 
well-known methods in statistical mechanics that 
\begin{equation} \label{e.Wclas}
W = N!{\prod_v}^* \frac{(dv/V)^{N_v \, dv}} {(N_v \, dv)!}
= \frac{N!} {V^N} {\prod_v}^* \frac{(dv)^{N_v \, dv}} {(N_v \, dv)!}\,.
\end{equation}
The asterisk means that we must multiply over groups of single  
particle states instead of all states, according to the cell division.
In the last term on r.h.s. of Eq. (\ref{e.Wclas}), $V$ was removed from 
the product taking into account that $\sum_{v}^* N_v dv = N$. 
We stress that the assignation of a volume $v$ to each 
particle does not imply a knowledge of the particle positions
in the physical space or that the particles should be at specific sites. 
In this sense, the present configurational description of a system of 
particles can be applied to different matter structures (gases, liquids 
and solids).

The calculation of $S_{conf}$ is straightforward. Using 
Stirling's theorem in the form $m! \approx (m/e)^m$, we obtain 
\begin{equation} \label{e.entropy}
S_{conf} = -k\int_0^V N_v \ln \left( \frac {N_v V} N \right) dv \,,
\end{equation}
where we have returned to an enumeration over all one-particle states instead 
of groups of states and replaced a sum over $v$ by an integral.

{\em The Helmholtz energy.}
The free energy of the gas is the sum of three terms
\begin{equation} \label{e.Helm}
F = F_{trans} +  F_{conf} + U_{int} \,. 
\end{equation}
$F_{conf}=-TS_{conf}$ is the configurational free-energy (being $T$ the
temperature of the gas) and $F_{trans}$ the translational free-energy, which
in equilibrium is given by, 
\begin{equation} \label{e.Ftrans}
F_{trans} = \frac N \beta \ln \left( \frac {\lambda^D} V \right) \,,
\end{equation}
where $\beta = 1/(kT)$ and $\lambda = (2\pi \hbar ^2 \beta /m)^{1/2}$ is the 
thermal de Broglie wavelength 
($\hbar=h/(2\pi)$ and $m$ is the mass of one particle).
$U_{int}$ is the sum of all energy contributions originated from the 
interactions among the particles. It is calculated as follows.

{\em Interaction energy.}
The average of the sum of two-particle interactions may be expressed as
\begin{equation} \label{e.Uint}
U_{int} = \frac{N}{2V} \int_0^V N_v \phi_v \, dv \,,
\end{equation}
where
\begin{equation} \label{e.phi}
\phi_v = \int_0^{[V]} u_v(\omega) \; g_v(\omega) \, d\omega \,.
\end{equation}
Here $u_v(\omega)$ is the pair interaction potential and $g_v(\omega)$ 
the pair distribution function (p.d.f.) corresponding to particles with 
available volume $v$. Both functions are expressed as a function of the volume 
$\omega$ enclosed by the interaction distance $r$.

\begin{figure}
\scalebox{0.43}{\includegraphics{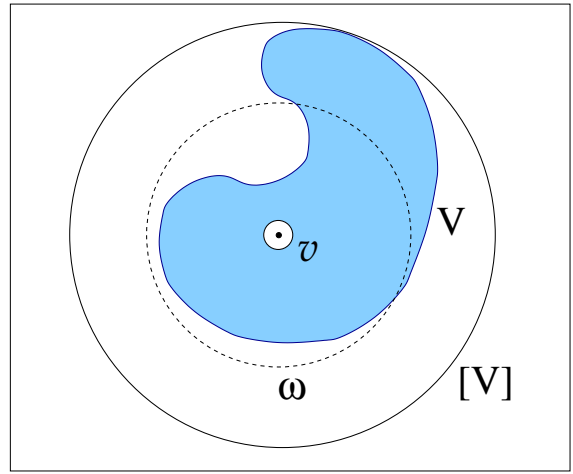}}
\caption{\label{f.vol1} Volumes involved in the evaluation of interaction
energy between a reference particle (central point) with available volume 
$v$ (schematically represented by a circle) and the rest of the gas contained 
in $V$ (shaded area). The volume $\omega$ centred in the chosen particle
is used to evaluated the interaction of this with the rest of the gas.
$[V]$ is the minimum volume $\omega$ which encloses the gas.} 
\end{figure}

Eqs. (\ref{e.Uint}, \ref{e.phi}) arise from considering the interactions
between a particle having an available volume $v$ and $n g_v(\omega) d\omega$ 
particles ($n=N/V$) located between the surfaces of spherical volumes $\omega$ 
and $\omega + d\omega$ centred on the chosen particle. 
The integration over $\omega$ takes into account all possible interactions 
of this particle with the remaining ones in $V$.
This evaluation is repeated (integration over $v$) for considering the 
interaction of each particle with the remaining gas.  
A factor of $1/2$ avoids the double count of binary interactions.
The upper limit $[V]$ of integration in Eq. (\ref{e.phi}) depends on the 
position of the reference particle within the gas volume and must be chosen 
as such that it completely covers $V$ (see Fig. \ref{f.vol1}).
However, it is assumed in Eq. (\ref{e.phi}) that $\phi_v$ is independent from 
the particle position in $V$ [since $u_v(\omega) \rightarrow 0$ 
rapidly as $\omega \rightarrow \infty$, for typical fluids].
We must adopt the convention that portions of $\omega$ outside the 
volume $V$ do not contribute to the integral. 
This prescription is applied to all integrals over $\omega$ of the present 
paper.
Since $n g_v(\omega)$ integrated over $V$ gives $N-1$
(i.e., the total number of particles around a chosen particle with available 
volume $v$), the p.d.f. $g_v(\omega)$ must verify
\begin{equation} \label{e.gcond}
\int_0^{[V]} g_v(\omega) d\omega = V- \frac 1 n \;.
\end{equation}

The dependence of the pair particle potential with the available volume 
assumed in Eq. (\ref{e.phi}) is not superfluous. It is well-known that 
many-body effects can modify the interaction between two particles. For 
example, $N$-body effects commonly soften the repulsive part of the two-body 
interactions in atomic and molecular fluids. 
In liquid hydrogen, the repulsion between two H$_2$ molecule 
is modified when a third molecule (or more) is placed nearby, because this 
induces changes in the electronic clouds of the particles \cite{ross}. 
A softening effect on pair potentials is also observed in liquid helium  
at high density \cite{aparicio}.
The confinement effects in the forces experimented by the particles can be
implicitly included in effective pair potentials (or so-called 
pseudopotentials) \cite{torrens}, for example with the adoption of a
density-dependent pair potential \cite{aparicio}. 
Alternatively, many-body effects in binary interactions could be represented 
in the present framework with a dependence of the interaction potential ($u_v$)
of a particle on the free space around it. It is worth noting that if the 
pair potential is independent from $v$, the interaction energy ($\phi_v$) of 
a particle can be even a function of the available volume throught the p.d.f.
($g_v$).

When the pair potential is independent from the available volume of the 
reference particle, Eqs. (\ref{e.Uint}-\ref{e.phi}) are reduced to
\begin{equation} \label{e.Uints}
U_{int} = \frac{N^2}{2V} \int_0^{[V]} u(\omega)\, g(\omega) \, d\omega \,,
\end{equation}
where 
\begin{equation} \label{e.gmedio}
g(\omega) \equiv \langle g_v(\omega) \rangle = \frac 1 N \int_0^V g_v(\omega) 
\, N_v \, dv \;.
\end{equation}
$g(\omega)d\omega$ is the probability to find a particle at a distance between 
$r$ and $r+dr$ ($\omega$ being the size of a $D$-sphere of radius $r$)
from another fixed particle regardless of its available volume. 
Therefore, this p.d.f., written in terms of the interparticle distance
$r$, coincides with the radial distribution function $g(r)$
typically used in theories of liquids and plasmas (e.g., \cite{standard}). 
Consequently, Eq. (\ref{e.Uints}) is the standard expression of 
the interaction energy in a fluid.
Therefore, the conventional form of the total potential energy of a 
one-component, classical fluid is automaticaly recovered from Eqs. 
(\ref{e.Uint}-\ref{e.phi}), as well as the p.d.f. $g(r)$
is recovered from Eq. (\ref{e.gmedio}) using the more detailed p.d.f. 
$g_v(\omega)$.

\section{ Equilibrium States} \label{s.equ_sta}

The three contributions to the r.h.s. of Eq. (\ref{e.Helm}) may be joined
together into a unique expression. The total free-energy of the gas in 
kinetic equilibrium now reads
\begin{equation} \label{e.F}
F = \int_0^V N_v \left[ \frac 1 \beta \ln \left( \frac {N_v \lambda^D} N 
\right) + \frac {N \phi_v}{2V} \right] dv \,.
\end{equation}

At low densities, high-order correlations in the particle distribution may 
be neglected so that $g_v(\omega)$ and (therefore) $\phi_v$ are independent
from $N_v$. The exact form of $g_v(\omega)$ depends on the considered gas. 
We shall return to this point later (Secs. \ref{s.ideal} and \ref{s.hard}). 

The equilibrium of the particle distribution is determined by the usual method.
For low density gases, the minimization of $F$ with the subsidiary 
conditions [Eqs. (\ref{e.Ncons}-\ref{e.Vcons})] yields
\begin{equation} \label{e.variat}
\int_0^V \delta N_v \left[ \frac 1 \beta \ln \frac{N_v \lambda^D }{ N }
+ \frac{N \phi_v}{2V} +\frac{U_{int}}{N} -\alpha 
+ \gamma v \right] dv = 0\;,
\end{equation}
where $\alpha$ and $\gamma$ are the Lagrange's parameters associated with
conservations of the particle number and the volume.
Since Eq. (\ref{e.variat}) must be verified for arbitrary functional variations 
$\{ \delta N_v \}$, we conclude that the equilibrium 
density is
\begin{equation} \label{e.Nv}
N_v = \frac{N}{\lambda^D} \exp \left[ -\beta \left( \gamma v + 
\frac{N \phi_v}{2V} +\frac{U_{int}}{N} - \alpha \right) \right] \;.
\end{equation}
Substitution of this expression in Eq. (\ref{e.F}) gives the equilibrium  
Helmholtz energy, which can be easily written in the form of a
Euler-like equation
\begin{equation} \label{e.Fequil}
F = -\gamma V + \alpha N - U_{int} \;.
\end{equation}

The parameters $\gamma$ and $\alpha$ are closely related to the pressure $P$
and the chemical potential $\mu$. In order to demonstrate it, we first derive
two useful identities. 
By introducing Eq. (\ref{e.Nv}) in Eq. (\ref{e.Ncons}) we find that, 
\begin{equation} \label{e.star1}
\lambda^D = \int^V_0 A dv \;,
\end{equation}
where $A$ is the exponential term of $N_v$ in Eq. (\ref{e.Nv}). 
Besides, since the current 
independent variables are the temperature, the particle number and the volume, 
we formally have $\alpha = \alpha(T,N,V)$ and $\gamma = \gamma(T,N,V)$.
Differentiating expression (\ref{e.star1}) with respect to $N$ (using 
the Leibniz's theorem \cite{Abra}) 
we obtain
\begin{eqnarray} \label{e.ident1a}
\left( \partial_N \alpha - \frac{\partial_N U_{int}}{N} + \frac{U_{int}}{N^2} 
\right) \int^V_0 A dv  \cr - \partial_N \gamma \int^V_0 v A dv -\frac{1}{2V}
\int^V_0 \phi_v A dv = 0 \,.
\end{eqnarray}
Integrals in Eq. (\ref{e.ident1a}) may be written in terms of $N$, $V$ and 
$U_{int}$. A little algebra leads to the first identity, 
\begin{equation} \label{e.ident1}
N \partial_N \alpha - V \partial_N \gamma - \partial_N U_{int} = 0 \;.
\end{equation}
The second one follows from the differentiation of Eq. (\ref{e.star1}) with 
respect to $V$. The result is
\begin{eqnarray} \label{e.star2}
N \partial_V \alpha - V \partial_V \gamma - \partial_V U_{int} 
+ \frac{U_{int}}{V} \cr 
- \frac{N}{2V} \int^V_0 \partial_V \phi_v N_v dv = 0 \;.
\end{eqnarray}
Here a term proportional to $N_{v=V}$ has been dropped since it tends to zero
in the thermodynamic limit (T-limit, $N,V \rightarrow \infty$ keeping $N/V$ 
constant). Typically, the interaction potential $u_v$ does not depend on $V$, 
so the integral term in Eq. (\ref{e.star2}) may also be neglected in the 
T-limit. Then, the second identity takes the form of
\begin{equation} \label{e.ident2}
N \partial_V \alpha - V \partial_V \gamma - \partial_V U_{int} 
+ \frac{U_{int}}{V} = 0 \;.
\end{equation}

From the usual thermodynamic relations and Eq. (\ref{e.Fequil}) it follows 
that
\begin{equation}
\mu \equiv \partial_N F = \alpha + N \partial_N \alpha - V \partial_N \gamma 
- \partial_N U_{int} \;,
\end{equation}
\begin{equation}
P \equiv -\partial_V F = \gamma + V \partial_V \gamma - N \partial_V \alpha
+ \partial_V U_{int} \;.
\end{equation}
With the help of identities (\ref{e.ident1}) and (\ref{e.ident2}), we find
that,
\begin{equation} \label{e.alphamu}
\mu = \alpha \;,
\end{equation}
\begin{equation} \label{e.gammaP}
P = \gamma + \frac {U_{int}}{V} \;.
\end{equation}
Therefore, $\alpha$ is identified with the chemical potential, while 
$\gamma$ equals to the pressure minus the density of interaction energy.
Substituting Eqs. (\ref{e.alphamu}) and (\ref{e.gammaP}) into Eq. 
(\ref{e.Fequil}) yields
\begin{equation} \label{e.euler}
F = -P V + \mu N  \;,
\end{equation}
according to the Euler's relation \cite{Callen}.

The program for the application of the present theory is as follows.
Given the pair potential of a fluid, the evaluation of Eqs. (\ref{e.Ncons})
and (\ref{e.Vcons}) with the equilibrium distribution [Eq. (\ref{e.Nv})],
yields the chemical potential $\mu$ and $\gamma$ in terms of $T$, $N$, and $V$.
Substitution of these solutions and $U_{int}$ [which is a function of 
$N$ and $V$, see Eq. (\ref{e.Uint})] into Eq. (\ref{e.Fequil}) gives
the characteristic function $F = F(T,N,V)$.
As it is well-known, the availability of the thermodynamic potential $F$ in
terms of the independent variables $T$, $N$ and $V$, provides us with the full
thermodynamics description of the system.

\section{Perfect Gas} \label{s.ideal}

For a gas composed of non-interacting particles
$u_v(\omega) = \phi_v = U_{int} \equiv 0$. The chemical  
potential $\mu$, $\gamma$ and the space distribution are determined
by substituting Eq. (\ref{e.Nv}) into the conditions given in Eqs. 
(\ref{e.Ncons}) and (\ref{e.Vcons}). Hence we obtain, 
\begin{equation} \label{e.mu_id}
\mu= kT \ln \left( \frac{\lambda^D \beta \gamma}{1-e^{-\beta \gamma V}} \right)
\,,
\end{equation}
\begin{equation} \label{e.gamma_id}
\beta \gamma=\frac N V \left[ \frac{1- ( 1+ \beta \gamma V) 
e^{-\beta \gamma V}} {1-e^{-\beta \gamma V}} \right] \,,
\end{equation}
\begin{equation} \label{e.N_vsimple}
N_{v}=N\frac{\beta \gamma \,e^{-\beta \gamma v}}{1-e^{-\beta \gamma \, V}} \,.
\end{equation}
Since there is no interaction energy in the gas, $P=\gamma$.
In the T-limit, equations of state (\ref{e.mu_id}) and (\ref{e.gamma_id}) 
recover their usual forms ($\beta \gamma \rightarrow n$)
\begin{eqnarray} \label{e.g_id}
\mu= kT \ln \left( n \lambda^D \right) \,, \cr
P = nkT \hspace{0.49in}\,, 
\end{eqnarray}
while the spatial distribution of particles is described by 
\begin{equation} \label{e.ddens}
n_v \equiv \frac {N_v} V = n^2 \exp \left( -nv \right) \,.
\end{equation}
Henceforth, $n_v$ is called {\em double density} because its units.
From Eqs. (\ref{e.euler}) and (\ref{e.g_id}), the free energy takes the 
expected form $F = NkT \left[ \ln \left( n \lambda^D \right) -1 \right]$.

\subsection{Identification of $v$} \label{ss.v}

We have noted in \S \ref{s.sp_st} that the elucidation of the available volume 
per particle is non-trivial and difficult to find previous to the 
application of the equilibrium theory. The present application to a perfect gas
shows that the volume $V$ is distributed among the particles following a 
decreasing exponential law [Eqs. (\ref{e.N_vsimple}) and (\ref{e.ddens})]. 
Most particles have available volumes going to zero, while few particles have 
a larger self space. This behavior is the same for all dimensions $D=1$, 2
and 3.

It is instructive to compare this result with the size distribution of 
Vorono\"{\i} regions associated to systems of non-interacting particles.
The size distribution of Vorono\"{\i} tesselations in one 
dimension is given exactly by a $\Gamma$-distribution \cite{kiang}
\begin{equation} \label{e.kiang}
n_v^{(Voronoi)} = \frac {cn^2}{\Gamma(c)} (cnv)^{(c-1)}\exp(-c n v)  \,,
\end{equation}
where $c=2D$ and $\Gamma$ denotes the usual gamma function. 
For higher dimensions ($D>1$) there are no exact results, although the size 
distributions can be even fitted by Eq. (\ref{e.kiang}) for two and three
dimensions \cite{kiang}. 
As we mentioned in Sec. \ref{s.sp_st}, the Vorono\"{\i} 
regions consist of all points closest to the particle chosen and, therefore,
represent segments, poligons and polyhedra for $D=1$, 2 and 3, respectively. 
Althought the Vorono\"{\i} space partition could be considered a good 
cantidate for the equilibrium distribution $N_v$ derived in Sec. 
\ref{s.equ_sta} [Eq. (\ref{e.Nv})], 
by comparison of Eqs. (\ref{e.ddens}) and (\ref{e.kiang}),
it is clear that the available volume per particle adopted by the 
thermostatistical theory applied to perfect gases is not a Vorono\"{\i} region. 

\begin{figure}
\scalebox{0.36}{\includegraphics{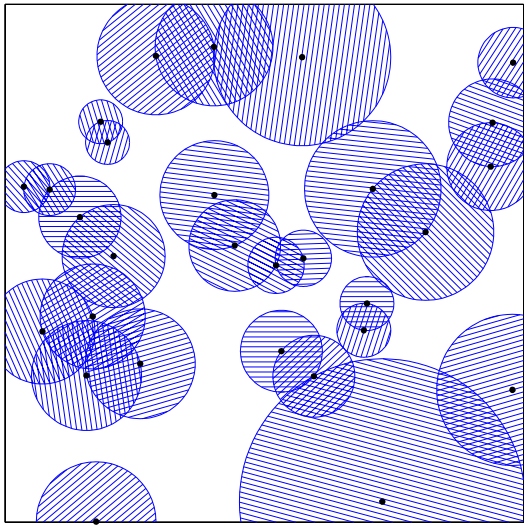}}
\caption{\label{f.circ} Typical configuration of uncorrelated points in two
dimensions. Shaded circles denote the available volumes of the objects
(see text).}
\end{figure}

However, there is a simple interpretation of $v$ for a perfect gas, as we
now observe. 
We notice that the distribution $n_v$ obtained for a system of 
non-interacting particles is equivalent (after a normalization) to the
Hertz or nearest-neighbor distribution function $H(r)$ associated to such 
systems \cite{chandra}.
In fact, the number $n_v dv$ of particles with available volume between $v$ 
and $v+dv$ given by Eq. (\ref{e.ddens}) is just the number $H(r)dr$ of 
particles which have the center of the 
nearest particle lying at a distance between $r$ and $r+dr$, i.e.,
between the surfaces of $D$-dimensional spheres of sizes $v(r)=v$ and 
$v(r+dr)=v+dv$ centred on the reference particle [see Eq. (\ref{e.Hrideal})].
This analogy suggests that the volume $v$ of a particle in a perfect gas in  
equilibrium can be directly related to its distance to the nearest neighbor.
Specifically, for a system of non-interacting bodies we propose the following 
identification of the available volume per particle,
\begin{equation} \label{e.def_vi}
\begin{array}{c}
v \equiv D-$dimensional sphere of radius given by the$  \\ 
\hspace{.23in} $distance between the centers of the reference$ \\
$particle and its nearest neighbor.\hspace{0.49in}$ 
\end{array}
\end{equation}
Fig. \ref{f.circ} illustrates the available volumes associated to a 
two-dimentional collection of Poisson points.
One notices that the proposal (\ref{e.def_vi}) does not establish an 
ownership connection between points of the space and each particle, 
so that the space is not divided into distinct, non-overlapping regions.
Nevertheless, we have not found any conflict between this identification of 
$v$ and the theory. 
On the contrary, this proposal for $v$ contains two useful advantages. 
First, as we shall show below, the present identification of $v$ leads to a 
simple and accurate representation of the p.d.f. $g_v$.
Second, $v$ gives an adequated notion of the free-particle space surrounding 
each particle and provides an useful tool for evaluating the perturbations
experimented by atoms and molecules in diluted real gases. We will return to
this point in \S \ref{s.atoms}.

\begin{figure}
\scalebox{0.43}{\includegraphics{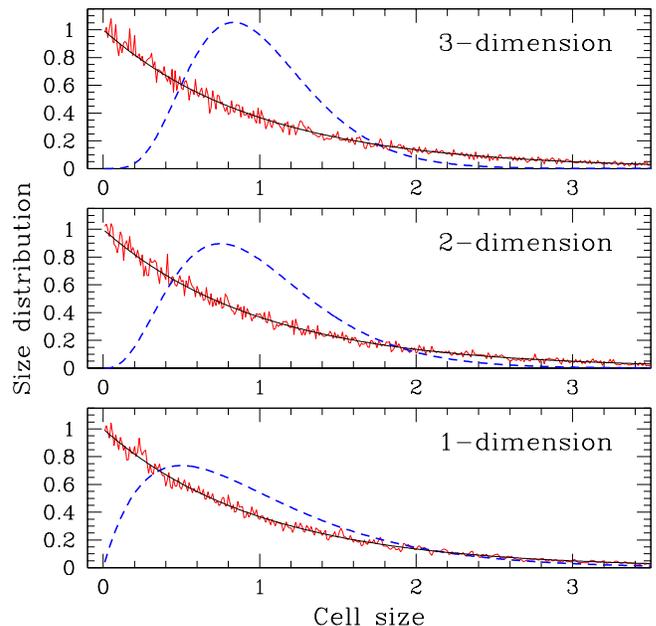}}
\caption{\label{f.w123} Results obtained from a simulation with 
$5 \times 10^4$ Poissonian points in one, two and three dimensions.
Exponential smooth curves correspond to the theoretical prediction
[Eq. (\ref{e.ddens})].
The Vorono\"{\i} statistics given by Eq. (\ref{e.kiang}) is shown by 
comparison (dashed lines).}
\end{figure}

Fig. \ref{f.w123} shows a direct test of Eq. (\ref{e.ddens}) and the 
identification (\ref{e.def_vi}) by computer simulation. 
Systems of $N=10^5$ uncorrelated points were simulated in unitary spaces 
of one, two and three dimensions using a standard algorithm \cite{press}. 
The exponential decreasing behavior of the size distribution of available
volumes is clearly reproduced by the numerical experiments. As a reference, 
one can observe in the figure the notably different laws followed
by the Vorono\"{\i}'s space partitions (dashed lines).

\subsection{Pair distributions} \label{ss.pd}

\begin{figure}
\scalebox{0.43}{\includegraphics{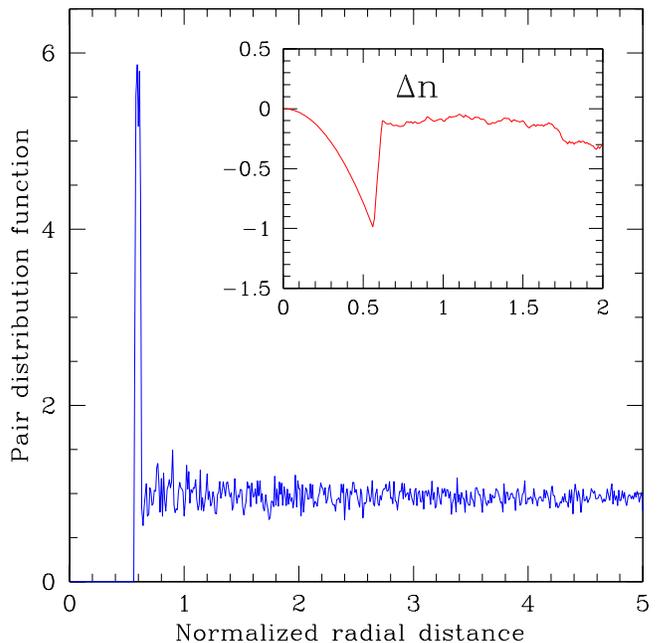}}
\caption{\label{f.g_2d} Pair distribution function $g_v$ for particles with
available volume $v=(1.1 \pm 0.1) n^{-1}$ in a two-dimension ideal gas as
a function of the dimensionless distance $x=r \sqrt n$ (units of the mean 
distance between particles). The peack around $x= 0.59$ is the contribution 
of the nearest neighbor.  
The results are for $N=10000$ particles randomly distributed in a unitary 
square surface (662 particles were selected with the chosen $v$).
The insert graph shows $\Delta n =\int_0^x (g_v-1)\pi x dx$, i.e., the 
difference between the actual particle number within a radius $x$ with respect 
to that one corresponding to a (microscopical) uniform particle density. 
The unitary jump (within the numerical errors) in $x= 0.59$ confirms that the 
peak of $g_v$ corresponds to {\em one} particle (the nearest neighbor).}
\end{figure}

From the identification (\ref{e.def_vi}), the evaluation of 
the p.d.f. $g_v(\omega)$ associated to particles with available volume $v$ 
is immediate. We notice that $v$ gives explicitly the distance 
to the first neighbor, which has a contribution $g_v^{(1)}$ to $g_v$.
Since $g_v^{(1)}(\omega) d\omega$ represents the probability of finding the 
nearest neighbor between the surfaces of spheres $\omega$ and $\omega+d\omega$,
one find that $n \int_0^\omega g_v^{(1)}(\omega') d\omega'$ must be zero 
for $\omega<v$ and one otherwise. Therefore, 
\begin{equation}
g_v^{(1)}(\omega) = \frac 1 n \delta(\omega -v) \,,
\end{equation}
where $\delta (\omega -v)$ is the Dirac delta function. 
On the other hand, $v$ does not give any information about the positions of 
the remaining particles in the gas except that they are outside of a spherical 
volume $v$ centred on the reference particle. Besides, these neighbors
are uncorrelated with the reference particle. Thus, their total contribution 
to $g_v$ must be one for $\omega>v$ and zero in the contrary case, which 
constitutes a Heaviside step function $\Theta (\omega-v)$.
Therefore, the p.d.f. of reference particles with available volume $v$ 
can be expressed in the form
\begin{equation}\label{e.gv_ideal}
g_v (\omega) = \frac 1 n \delta(\omega -v) + \Theta (\omega-v) \,.
\end{equation}
Fig. \ref{f.g_2d} confirms our identification of $g_v(\omega)$. 
It is also worthwhile to notice that Eq. (\ref{e.gv_ideal}) reproduces the 
expected value of the averaged p.d.f. [Eq. (\ref{e.gmedio})] for a set 
of non-interacting particles,
\begin{equation}
g (\omega) = \int_0^\infty g_v(\omega) n e^{-nv} dv
= e^{-n\omega} + 1 - e^{-n\omega} = 1 \,,
\end{equation}
which expresses that the positions of the particles are completly 
uncorrelated and, therefore, the probability density of finding neighbors 
(without any specification of the available volume of the reference particle) 
is uniform and equals unity. We notice also that 
$g^{(1)}(\omega) = \langle g_v^{(1)}(\omega) \rangle =e^{-n\omega}$ 
represents the first neighbor distribution for random configurations 
of non-interacting points. The nearest-neighbor distribution function $H(r)$,
which was introduced in \S \ref{ss.v}, can be calculated from the relation
\begin{equation} \label{e.Hr}
H(r)=n \frac{d\omega(r)}{dr} g^{(1)}\left[\omega(r)\right] \,.
\end{equation}
Its evaluation reproduces the well-known results of a perfect gas 
\cite{torquato95},
\begin{equation} \label{e.Hrideal}
H(r)=n \frac{d\omega(r)}{dr} \exp \left[ -n\omega(r) \right] \,.
\end{equation}
For $D=3$, it is a classical result derived by Hertz \cite{hertz}.

In closing this section, it is worth stressing that all termodynamics and 
spatial 
statistical, well-known results of classical perfect gases in one, two and 
three dimensions, are self-consistently derived from the theory developed in 
Secs.  \ref{s.ther} and \ref{s.equ_sta}. 
The identification of the available volume given in (\ref{e.def_vi}) plays
a central role in the present application of the theory, and so we believe 
that it should be regarded as the most plausible interpretation of $v$ for 
systems of non-interacting particles.

\section{Gases of Hard Elastic Particles} \label{s.hard}

Systems formed by hard bodies constitute simple fluid models in which 
impenetrable particles interact solely via hard-core repulsions.
These models have played a major role in the liquid state theory,
as the behavior of dense fluids is dominated by the exclude-volume effects 
associated with the repulsive forces of its constituents.
The main advantage of these models is the simplicity of the pair potential, 
which can be defined by
\begin{equation} \label{e.uv_hs}
u_v(\omega) = \left\{ 
\begin{array}{c}

  \infty,      \hspace{.43in}  \omega \le a \\  
  0,           \hspace{.52in}  \omega > a
\end{array}
\right.
\end{equation}
where $a$ is the size of a $D$-sphere of radius $d$, $d$ being the diameter
of a particle.
The aim of this section is to apply the theory to hard-particle systems,
which can provide a guidance for the treatment of more realistic fluids.
We concentrate on the limit of low densities, although the results derived
for $D=1$ will be valid at all densities.
\begin{figure}
\scalebox{0.30}{\includegraphics{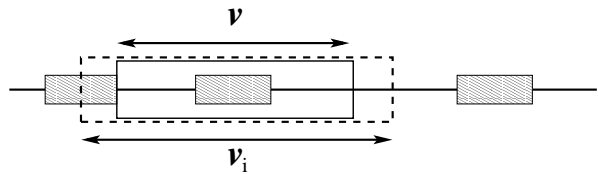}}
\caption{\label{f.rods} Schematic representation of the available volume
$v$ of a hard rod (central shaded rectangle) in presence of its nearest
neighbor. The volume $v_i$ is the (``ideal'') available volume of the central 
body if the bodies are fully penetrable (non-interacting), 
according to the interpretation (\ref{e.def_vi}).} 
\end{figure}

As a consequence of the hard-core interaction, the identification 
(\ref{e.def_vi}) of the available volume of a particle is not valid for the 
present fluid.
In fact, we notice that the space occupied by a particle is inaccesible to the 
other ones, therefore this region must be a part of the available volume of 
the particle not shared with the others. Consequently, the space occupied 
by neighbor particles yields a reduction of the available volume of a particle 
with respect to that ($v_i$) calculated for fully penetrable (non-interacting) 
objects, so that $v<v_i$. As an {\em ansatz}, we propose that the available 
volume of a hard $D$-sphere is given by
\begin{equation} \label{e.vid}
v = v_{i} - a^* \,,
\end{equation}
where $a^*$ is the reduction of the available volume of a hard particle with 
respect to the perfect-gas value (\ref{e.def_vi}). As we shall show 
later, the value of $a^*$ can be evaluated using the virial theorem.
\begin{figure}
\scalebox{0.38}{\includegraphics{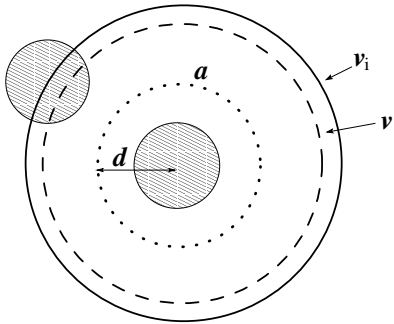}}
\caption{ \label{f.vid_hs} Same as Fig. \ref{f.rods}, but for $D=2$. 
The exclusion region $a$ (circle closed by dotted points) is a forbidden 
region to the particle center of neighbors.
The available volume of a hard particle ($v$) is lower than the perfect-gas 
value ($v_i$) (see text).} 
\end{figure}

On the other hand, one expects that the p.d.f. $g_v$ of a perfect gas 
[Eq. (\ref{e.gv_ideal})] is a good approximation to the p.d.f. of 
non-ideal fluids at very low densities. If we take into account the relation 
(\ref{e.vid}), the p.d.f. of a diluted hard-body fluid can be written as
\begin{equation}\label{e.ghard}
g_v (\omega) = \frac 1 n \delta(\omega -v_{i}) + \Theta (\omega-v_{i}) \,.
\end{equation}
From Eqs. (\ref{e.phi}), (\ref{e.uv_hs}) and (\ref{e.ghard}), we find the 
mean potential energy of a particle with available volume $v$,
\begin{equation} \label{e.phi_hs}
\phi_v = \left\{ 
\begin{array}{c}
  \infty,      \hspace{.43in}  v \le b \\  
  0,           \hspace{.52in}  v > b
\end{array}
\right.
\end{equation}
where
\begin{equation}\label{e.baa}
b = a - a^* \,.
\end{equation}
From Eq. (\ref{e.Nv}) then results that $N_v$ is zero for $v \le b$ and has an 
exponential dependence on the available volume for $v>b$. 
Clearly $U_{int}=0$ and then $P= \gamma$.
At the T-limit, evaluations of Eqs. (\ref{e.Ncons}-\ref{e.Vcons}) yield
\begin{equation} \label{e.a_hs}
\mu = kT \left[ \frac {nb} {1-nb} + \ln \left( \frac{n \lambda^D}{1-nb} 
\right) \right] \,,
\end{equation}
\begin{equation} \label{e.g_hs}
\gamma = \frac {n kT} {1-nb} \,,
\end{equation} 
\begin{equation} \label{e.nv_hs}
n_v = \Theta(v-b) \left( \frac{n^2}{1-nb} \right) 
\exp \left[ - \frac {n(v-b)}{1-nb} \right] \;.
\end{equation}
The free energy is easily derived from Eqs. (\ref{e.Fequil}), (\ref{e.a_hs}) 
and (\ref{e.g_hs}), and reads
\begin{equation} \label{e.F_hs}
F = NkT \left\{ \ln \left[ (n\lambda^D)/(1-nb) \right] -1 \right\}. 
\end{equation}

As we anticipated, the virial theorem may provide the value
of $b$ and, thus, the elucidation of the factor $a^*$ at low densities.
According to Eq. (\ref{e.g_hs}), $b$ represents the second virial coefficient 
in the equation of state $P=nkT(1+bn+...)$ (since $\gamma=P$) and, therefore,
its value is well-known for hard $D$-spheres. 
If we denote with $\sigma$ the size of a hard-particle, then $b=\sigma$, 
$2\sigma$ and $4\sigma$, for $D=1$, 2 and 3 \cite{santos}. 
Therefore, from Eq. (\ref{e.baa}) and the known values of $a$, it follows that
$a^* = b = a/2$. 
Explicit values of interest are given in Table \ref{t.table}.
\begin{table}
\caption{\label{t.table} Quantities used in the study of 
$D$-dimensional hard-body fluids. 
$\omega(r)$, size of a $D$-sphere of radius $r$; 
$\sigma$, volume of a hard-body of diameter $d$; $\eta$, packing fraction; 
$a$, exclusion sphere of the repulsive interaction; $a^*$, reduction of the 
available volume per hard-body; $b=a-a^*$ (see text).}
\begin{ruledtabular}
\begin{tabular}{lrrrrrr}
   & $\omega(r)$ & $\sigma$ & $\eta$ & $a$ & $a^*$ & $b$ \\
\hline
$D=1$ & $2r$      &  $d$ & $n\sigma$   & $2\sigma$ &  $\sigma$ &  $\sigma$ \\   
$D=2$ & $\pi r^2$ & $\pi d^2/4$ & $n\sigma$ & $4\sigma$ & $2\sigma$ & 
$2\sigma$ \\       
$D=3$ & $4\pi r^3/3$ & $\pi d^3/6$& $n\sigma$& $8\sigma$& $4\sigma$ & 
$4\sigma$    \\  
\end{tabular}
\end{ruledtabular}
\end{table}

The averaged nearest-neighbor p.d.f. can be calculated substituting  
the first term on the r.h.s. in Eq. (\ref{e.ghard}) and $n_v$ given by Eq. 
(\ref{e.nv_hs}) into Eq. (\ref{e.gmedio}). Thus, we obtain that, 
\begin{equation} \label{e.g1_hs}
g^{(1)}(\omega) = \Theta(\omega-a) \left( \frac{1}{1-nb} \right) 
\exp \left[ - \frac {n(\omega-a)}{1-nb} \right] \;.
\end{equation}
Further, the p.d.f. for all neighbors except the nearest one can be derived 
in similar way from the second term on the r.h.s. of Eq. (\ref{e.ghard}).
The result is
\begin{equation} \label{e.gr_hs}
g^{(r)}(\omega) = \Theta(\omega-a) \left\{ 1- \exp \left[ - 
\frac {n(\omega-a)}{1-nb} \right] \right\} \;.
\end{equation}
The total averaged p.d.f. is composed by the sum of these two contributions.

Knowledge of $g^{(1)}(\omega)$ permits us to evaluate the nearest-neighbor 
distribution function $H(r)$ from Eq. (\ref{e.Hr}). For convenience, 
we introduce the 
dimensionless distance $x=r/d$ and the packing fraction $\eta= n \sigma$, 
which gives the fraction of the total volume occupied by the hard bodies.
Then $H(x)= d\;H(r)$, and we obtain
\begin{equation} \label{e.H1_hs}
H(x) = \frac {2 \eta}{1-\eta} \exp \left[ \frac{-2\eta \left( x-1 \right) }
{1-\eta} \right] \;, \hspace{.35in}  x>1
\end{equation}
\begin{equation} \label{e.H2_hs}
H(x) = \frac {8 \eta x}{1-2\eta} \exp \left[ \frac{-4\eta \left(x^2-1 \right)}
{1-2\eta} \right] \;, \hspace{.20in}  x>1
\end{equation}
\begin{equation} \label{e.H3_hs}
H(x) = \frac {24 \eta x^2}{1-4\eta} \exp \left[ \frac{-8\eta \left( x^3-1 
\right) } {1-4\eta} \right] \;, \hspace{.20in}  x>1
\end{equation}
for $D=1$, 2 and 3, respectively ($H=0$ at $x \le 1$).

\subsection{Hard-rod fluid} \label{s.hrf}

The properties of a fluid of hard rods ($D=1$) are known exactly, hence
this model offers an appropriate test for our evaluations. Thus, it can 
be verified that the free energy given in Eq. (\ref{e.F_hs}) is rigorously 
exact for $D=1$ and, therefore, all the equations of state are correct.
In particular, Eq. (\ref{e.g_hs}) gives the exact equation of state for 
hard rods as derived earlier by Tonks \cite{tonks}. 

\begin{figure}
\scalebox{0.43}{\includegraphics{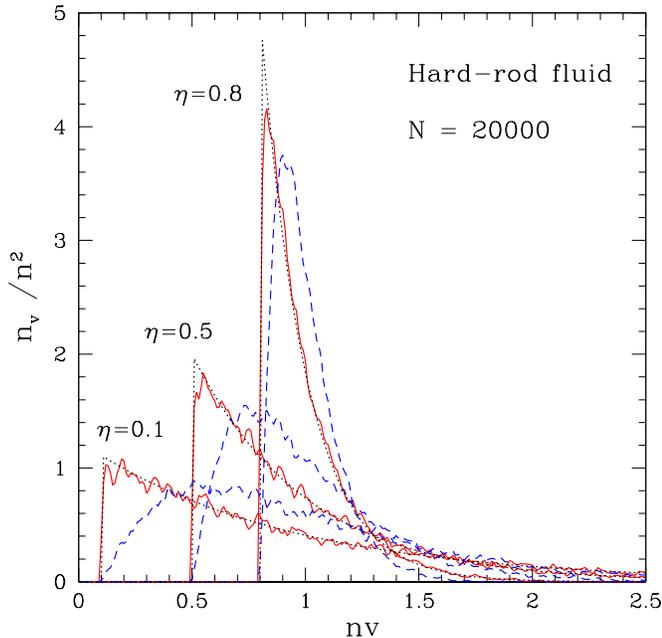}}
\caption{\label{f.nv158} Dimensionless double density of a hard-rod fluid 
for values of the packing fraction $\eta=n\sigma=0.1$, 0.5, and 0.8.
Dotted curves correspond to the theoretical prediction [Eq. (\ref{e.nv_hs})],
and solid lines represent the results obtained from numerical simulations with 
$2 \times 10^4$ hard particles. 
The Vorono\"{\i} statistics is shown by comparison (dashed lines).}
\end{figure}

As we have seen in the case of a perfect gas, the present theory not only 
gives the thermodynamic properties of the fluid, but also provides information 
on its microstructure throught the doble density $n_v$. 
Fig. \ref{f.nv158} shows $n_v$ for several values of the packing fraction 
$\eta=0.1$, 0.5 and 0.8, as obtained from Eq. (\ref{e.nv_hs}).
Due to the effect of the core repulsions [see Eq. (\ref{e.phi_hs})], no 
particle may have an available volume lower than $b$, so that the distribution 
$n_v$ is characterized by an abrupt fall in $v=b$ ($nv=\eta$ in the figure).
The distribution $n_v$ becomes more and more peaked as the gas concentration 
increases. This prediction is confirmed by numerical simulations
(Fig. \ref{f.nv158}). 

It may surprise us that the p.d.f. $g_v$ given by Eq. (\ref{e.ghard}) is only 
an approximated expression (its form was adopted from the perfect-gas solution),
 while its application leads to correct results as previously described.
This is due to the fact that the approximation made on $g_v$ rests on the 
second term on the r.h.s. in Eq. (\ref{e.ghard}) (it ignores pair correlations 
between a particle and its second, third, etc., neighbors), which is not 
required for the determination of the double density $n_v$. 
In a linear gas only the p.d.f. of the nearest neighbor is relevant, being
correctly given by the first term on the r.h.s. in Eq. (\ref{e.ghard}). 
Thus, Eq. (\ref{e.g1_hs}) is exact for $D=1$ and leads to the 
correct distribution $H(r)$ of the nearest neighbor, as given in Eq. 
(\ref{e.H1_hs}). This result was derived earlier by MacDonald \cite{macdonald}.

Finally, we recall that the present theory yields exact results for the
thermodynamic and geometric structure of hard rods. It gives an important 
support to the identifications (\ref{e.def_vi}) and (\ref{e.vid}) of the 
available volume per particle.

\subsection{Hard disks and spheres} \label{s.hds}

Not a single exact result is known for $D > 1$, but a number 
of numerical solutions and analytical approximations are available
\cite{baus_colot}. 
Inspection of the free energy [Eq. (\ref{e.F_hs})] and 
equations of state (\ref{e.a_hs}) and (\ref{e.g_hs}) evaluated with values of
$b$ given in Table \ref{t.table} reveals that our results are found to give 
the correct thermodynamical properties of hard discs and 
spheres at low densities.

\begin{figure}
\scalebox{0.43}{\includegraphics{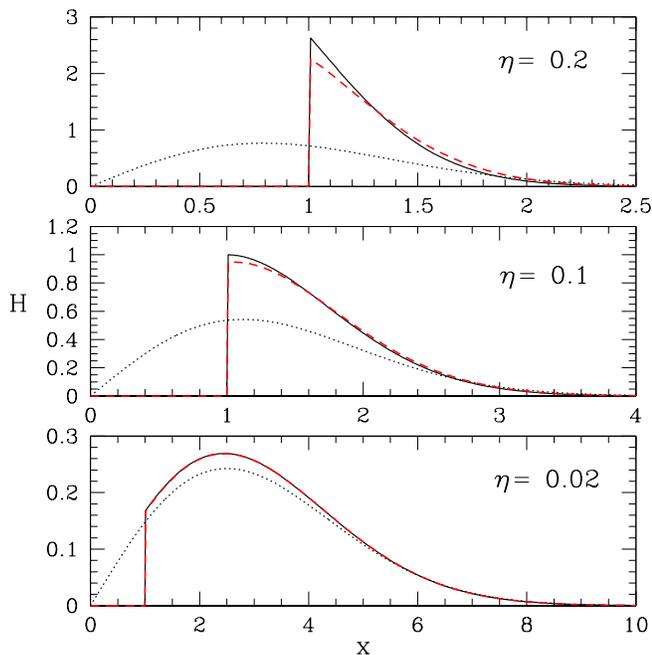}}
\caption{\label{f.HD2} Nearest neighbor function $H(x)$ of a hard-disc
fluid at packing fraction $\eta=0.02$, 0.1, and 0.2, in terms of the 
reduced distance $x=r/d$ ($d$, diameter of a particle). 
Solid curves result from Eq. (\ref{e.H2_hs}), dashed lines correspond to 
accurate fits up to freezing ($\eta < 0.69$) \cite{torquato95}, 
and dotted lines are results for fully penetrable particles.}
\end{figure}

\begin{figure}
\scalebox{0.43}{\includegraphics{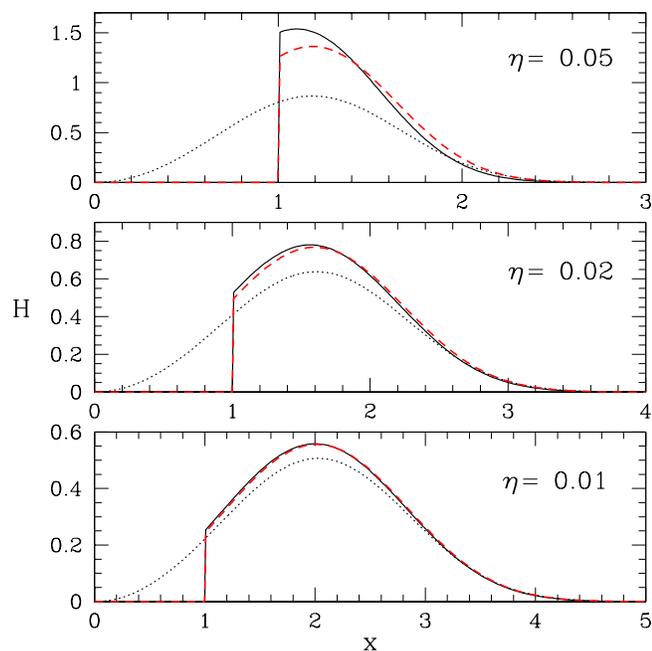}}
\caption{\label{f.HD3} Nearest neighbor function $H(x)$ of a hard-sphere 
fluid at packing fraction $\eta=0.01$, 0.02, and 0.05, in terms of the 
reduce distance $x=r/d$ ($d$, diameter of a particle). 
Solid curves result from Eq. (\ref{e.H3_hs}), dashed lines correspond to 
accurate fits up to freezing ($\eta < 0.49$) \cite{torquato95}, 
and dotted lines are results for fully penetrable particles.}
\end{figure}

In order to verify the double density $n_v$ given in Eq. (\ref{e.nv_hs}),
we study the nearest neighbor distribution function $H(x)$ which is
derived from it.
For this purpose, Eqs. (\ref{e.H2_hs}) and (\ref{e.H3_hs}) are contrasted 
with the analytical expressions derived by Torquato \cite{torquato95}, which 
show an excelent agreement with Monte Carlo simulation data \cite{torquatolee}.
We use Torquato's expressions valid for densities lower than 
the freezing density (the freezing packing fraction is $\eta_f \approx 0.69$ 
and $\eta_f \approx 0.49$ for discs and spheres, respectively).
Figs. \ref{f.HD2} and \ref{f.HD3} show that Eqs. (\ref{e.H2_hs}) and 
(\ref{e.H3_hs}) give the correct asymptotic behavior of $H(x)$ at low
densities, and capture the salient features of $H(x)$ for moderate densities.
The agreement between these results and Torquato's expressions is within  
1\% everywhere at packing fraction $\eta<0.01$ and $\eta<0.002$, 
for $D=2$ and $D=3$, respectively.

In closing, the low-density, asymptotic results of hard discs and spheres are 
correctly reproduced by the present theory. We find again a support for the
identification of the available volume per particle, closely related
with the distance to the nearest neighbor.

\subsection{Van der Waals fluid} \label{s.vdw}

In the preceding subsections, we have studied the application of the
theory to systems composed by particles with infinite hard cores. 
We shall consider now the previous model but with the 
inclusion of an attractive long-range contribution in the potential.
Because attractive forces do not restrict the accessible space to a 
particle, the identification (\ref{e.vid}) of the available volume per 
particle (with $a^*$ given in Table \ref{t.table}) should remain unaltered 
for this model.
Instead of trying to figure out the pair potential itself, one may directly
define $\phi_v$ as
\begin{equation} \label{e.phi_vdw} 
\phi_v = \left\{ 
\begin{array}{c}
  \infty,      \hspace{.45in}  v \le b \\  
  -\epsilon,   \hspace{.45in}  v > b
\end{array}
\right.
\end{equation}
where $\epsilon$ is a constant parameter (with units of energy times volume)
and $b$ is given by Eq. (\ref{e.baa}).
Eq. (\ref{e.phi_vdw}) is essential for obtaining the Van der Waals equation 
and is the simplest choice to test the relation (\ref{e.gammaP}) between the 
pressure and the parameter $\gamma$ when $U_{int}$ is non-zero. 

In fact, this form for $\phi_v$ is based on the suitable short-range repulsion 
and an attractive interaction whose range is finite but long compared to the 
interparticle spacing. An appropriate attractive potential with these 
characteristics is the general K\^ac potential \cite{Kac}
\begin{equation} \label{e.Kac}
u_v^{Kac} (\omega) = -\epsilon \gamma^* J (\gamma^* \omega)  \;,
\end{equation}
with $\int_0^\infty J(\omega) d\omega =1$. Here $\gamma^*$ represents 
the $D$th power of the usual inverse-range parameter.
The evaluation of the attractive-interaction energy per particle is immediate.
From Eqs. (\ref{e.phi}), (\ref{e.ghard}) and (\ref{e.Kac}), one obtains
(in the T-limit) that, 
\begin{equation} \label{e.phi_Kac}
\phi_v^{Kac} = -\frac {\epsilon}n \gamma^* J\left[\gamma^* (v+a^*)\right]
- \epsilon  \int_{\gamma^*(v+a^*)}^\infty J(\omega) \omega  \;.
\end{equation}
In the limit $\gamma^* \rightarrow 0$, $\phi_v^{Kac} \rightarrow -\epsilon$ 
and we recover the attractive term in Eq. (\ref{e.phi_vdw}). This result shows 
that the attractive energy of a particle results from its simultaneous 
interaction with a large number of particles, where the contribution of the
nearest neighbors is negligible regardless their distances to the reference 
one. As a consequence, $\phi_v^{Kac}$ is equivalent for all 
particles independently from their available volumes.

The interaction energy given in Eq. (\ref{e.phi_vdw}), after substitution in
Eq. (\ref{e.Nv}) and normalization, reproduces the double density given by 
Eq. (\ref{e.nv_hs}). 
In the T-limit, Eqs. (\ref{e.Ncons}) and (\ref{e.Vcons}) lead to
\begin{equation} \label{e.a_vdw}
\mu= kT \left[ \frac {nb} {1-nb} + \ln \left( \frac{n \lambda^D}{1-nb} 
\right) \right] - \epsilon n  \,,
\end{equation}
while expression (\ref{e.g_hs}) is obtained for $\gamma$.
The evaluation of Eq. (\ref{e.Uint}) yields a non-zero interaction energy with 
density $U_{int}/V = - \epsilon n^2 /2 $.
Besides, from previous results and Eq. (\ref{e.Fequil}), the Helmholtz energy 
takes the form of
\begin{equation} \label{e.f_vdw}
F = NkT \left\{ \ln \left[ n \lambda^D/(1-nb) \right] -1 \right\}
- \epsilon N n/2 \,.
\end{equation}
From differentation of $F$ with respect to $V$ one obtains the well-known 
Van der Waals equation of state,
\begin{equation} \label{e.P_vdw}
P = \frac {n kT} {1-nb} - \frac {\epsilon n^2} {2} \,.
\end{equation}
Relation (\ref{e.gammaP}) is thus confirmed beyond the trivial case 
($P=\gamma$) obtained for perfect and hard-body fluids.
The Van der Waals model constitutes a special case where attractive forces 
have not influenced the value of $\gamma$, which, however, is strongly 
governed by the infinitely repulsive interactions.

\subsection{Order in the fluid structure} \label{s.order}

We notice that the double density offers a way to measure the order of the 
packing of hard particles. In fact, we can define the fluctuation of the 
available volume as the standard deviation
\begin{equation} \label{e.Delta}
\Delta v \equiv \left[ \langle v^2 \rangle - \langle v \rangle ^2 \right]^{1/2}
\end{equation}
where $\langle ... \rangle$ represents an average in the ensemble 
[earlier applied in Eq. (\ref{e.gmedio})], namely, 
\begin{equation}
\langle f_v \rangle \equiv \frac 1 n \int_0^\infty f_v \, n_v \, dv \;,
\end{equation}
$f_v$ being a function of the available volume per particle. Obviously, 
$\langle v \rangle = n^{-1}$ for all fluids [see Eq. (\ref{e.Vcons})]. 
According to Eq. (\ref{e.nv_hs}) for hard-body fluids, one obtains
\begin{equation} \label{e.delta_hs}
\begin{array}{c}
\Delta v = \langle v \rangle  - b \;, \hspace{.24in} ($hard bodies$) \,.
\end{array}
\end{equation}
For comparison, the space fluctuation of classical perfect gases is as large
as the mean volume per particle for any temperature and density 
[c.f. Eq. (\ref{e.ddens})],
\begin{equation} \label{e.delta_id}
\begin{array}{c}
\Delta v= \langle v \rangle \;, \hspace{.53in}   ($perfect gas$) \,.
\end{array}
\end{equation}
This points out the irregular geometric structure that characterizes random
collections of non-interacting objects.

The volume fluctuation $\Delta v$ given by Eq. (\ref{e.delta_hs}), summarizes  
the behavior of $n_v$ with the density for hard rods. Large deviations of 
the available volumes from the mean value occur at low densities, and the 
hard-rod fluid has the inhomogeneity of a perfect gas in the limit 
$\eta \rightarrow 0$ ($\Delta v \rightarrow n^{-1}$). As expected, 
the fluctuations decrease
at high densities and vanish in the limit of close-packing $\eta=1$.
Eq. (\ref{e.delta_hs}) is valid for discs and spheres at low densities.

Classical methods for determining the order in an isotropic packing of hard 
$D$-spheres are based on geometrical and topological techniques.
The former rests on the identification of peaks in the radial 
function $g(r)$, which is found to be unsatisfactory because, among other 
reasons, it is difficult to determine when the peak appears \cite{rintoul}. 
Topological methods use the so-called order or topological parameters, such 
as like-coordination numbers \cite{steinhardt} and the number of faces in a 
cell division \cite{rivier, jullien}. 
They commonly require an appreciable amount of calculation divided in two 
steps, first, an evaluation of configurations of the system in equilibrium
states (usually based on numerical simulations), and, then, the execution 
of an algorithm which evaluates the cell division (typically a Vorono\"{\i} 
tesselation).

In constrast with the previous methods, the present formalism offers,
eventually, a direct and precise way to characterize the order in fluid 
structures in terms of the fluctuation of the available volume per particle. 
Thus, for example, $\Delta v = \langle v \rangle$ represents the high 
density fluctuations present in very diluted gases. 
On the contrary, uniformily ordered configurations should be characterized 
by space fluctuations noticeably lower than $\langle v \rangle$. 
Although the application of $\Delta v$ is here limited to hard particles 
at low densities, additional investigation could extend its use
to real fluids.

\section{Diluted hydrogen plasma} \label{s.atoms}

Finally, as the main example of this work we consider a low density gas 
formed by
hydrogen atoms in an electrically neutral background of free electrons and 
protons. 
For simplicity, we consider a strongly ionized equilibrium state where the 
atom-atom interaction can be ignored, and only the charge effects upon the 
atoms are taken into account. 
Each atom is considered as a polarizable particle and its available
volume is approximated by the perfect-gas value (\ref{e.def_vi}). 
Our aim focuses on the description of the neutral component of the gas.
We emphasize that the results obtained from this model are qualitative and 
illustrative only. A detailed study of a mixture of gases in equilibrium is in 
progress and its results will be presented elsewhere.

The equilibrium density given in Eq. (\ref{e.Nv}) can be easily generalized 
to particles with internal states following the procedure of Secs.
\ref{s.ther} and \ref{s.equ_sta}. The double density of atoms in the energy 
level $j$ and with available volume $v$ [hereafter, atoms ($j,v$)] is given by
\begin{equation} \label{e.Njv}
n_{j,v} = \frac {n}{ \lambda^3 }\, g_{j} \, \exp \left[-\beta \left( 
\epsilon_{j} + \Delta \epsilon_{jv} + \gamma v -\mu \right) 
\right] \;,
\end{equation}
where $g_{j}$ and $\epsilon_{j}$ are the multiplicity and eigenenergy
of the level $j$ for an isolated atom, $n$ is the total density,
and $\lambda$ and $\mu$ the thermal wavelength and chemical potential 
of the atomic component, respectively.
The factor $\Delta \epsilon_{jv}$ can be interpreted as the energy change of 
a particle ($j,v$) due to interactions with the charges. 
Specifically, we obtain
\begin{equation} \label{e.ejv}
\Delta \epsilon_{jv} = \frac 1 2 \left( n_e \phi_{jv,e} + n_p \phi_{jv,p}
\right) \;,
\end{equation}
with
\begin{equation} \label{e.phi_jv}
\phi_{jv,c} = \int_0^{[V]} u_{jv,c} (\omega)\; g_{jv,c}(\omega) \,d\omega \;,
\end{equation}
where the subscript $c$ refers to either electrons or protons, $n_e$ and $n_p$
are the number densities of free electrons and protons, $u_{jv, c}$ is the pair 
potential of the interaction of an atom ($j,v$) with a charge $c$,
and $g_{jv,c}(\omega)$ represents the probability density of finding 
a charge $c$ in the surface of a spherical volume $\omega$ centred on an
atom ($j,v$).  
We have omitted in Eq. (\ref{e.ejv}) the energy change of atoms 
in the level $j$ due to their perturbations {\em over} the charges, because
it is only a small correction (proportional to the abundance of atoms)
in $\Delta \epsilon_{jv}$ for the physical conditions assumed.

At large separations, charges induce a dipole moment in the hydrogen atom. 
The pair potential associated to polarization of an atom by either an 
electron or 
a proton can be represented by the screened Debye potential obtained by
Redmer and R\"opke \cite{redmer} (in the form used in \cite{potekhin})
\begin{equation} \label{e.pol}
u_{pol}(r) = -\frac{ e^2 \alpha_j } 2 \left( \frac {1+r/r_s}{l_j^2 + r^2}
\right)^2 e^{-2r/r_s}  \;,
\end{equation} 
where $e$ is the elemental charge, $r_s$ the screening length, $\alpha_j$ the
polarizability of an atom at the level $j$, and $l_j =a_B j[(7j^2+5)/4]^{1/2}$ 
(with $j$ the main quantum number and $a_B$ the Bohr radius) its mean radius.

At short separation between an atom and a proton the Coulomb repulsion 
between protons prevails \cite{potekhin}
\begin{equation} \label{e.rep_p}
u_{C}(r)= \frac {e^2} r \left( 1+ \frac r a \right) e^{-2r/a } \;.
\end{equation} 
This potential includes a plasma screening function with a 
characteristic length $a=\frac 12 a_{B} l_j/l_1$.

Due to Pauli's exclusion principle, there is a strong repulsion when the 
electronic wavefunctions of two atoms begin to overlap and this is often 
expressed as
\begin{equation} \label{e.rep_e}
u_{12}(r)= \frac {C_{12} } {r^{12}} \;.
\end{equation}
We use this potential to simulate the interaction between a free electron and 
an atom. The coefficient $C_{12}$ is harder to evaluate. From Hindmarsh et al.
\cite{hindmarsh} we adopt $C_{12}=9 \times 10^{-17} l_j$ erg cm$^{12}$, 
where we consider only the wavefunction size (of characteristic radius 
$l_j$) of the bounded electron.

On the average, roughly half of atoms have an electron as its nearest neighbor 
and the other half a proton. The fraction of atoms having other atoms as 
nearest neighbors is negligible in a highly ionized material. Therefore, the 
total density of atoms can be written as
\begin{equation} \label{e.nat}
n_H = \sum_j \int_0^\infty n_{jv}^{(p)} dv + \int_0^\infty n_{jv}^{(e)} dv \;,
\end{equation}
where $n_{jv}^{(p)}$ and $n_{jv}^{(e)}$ are the double densities of atoms which 
have a proton or an electron as their nearest neighbor. Both densities are 
given by Eq. (\ref{e.Njv}) (times $1/2$), although they differ in the energy 
term $\Delta \epsilon_{jv}$. For the first and second atom sets 
we have,
\begin{equation} \label{e.ejv_p}
\Delta \epsilon_{jv}^{(p)} = \frac 1 2 \left( u_p(r) + \int_v^\infty 
\left[ n_e u_e(\omega) + n_p u_p(\omega) \right] d\omega \right) ,
\end{equation}
\begin{equation} \label{e.ejv_e}
\Delta \epsilon_{jv}^{(e)} = \frac 1 2 \left( u_e(r) + \int_v^\infty 
\left[ n_e u_e(\omega) + n_p u_p(\omega) \right] d\omega \right) ,
\end{equation}
respectively,
where $u_p = u_{C} + u_{pol} $ and $u_e = u_{12} + u_{pol} $ are the total
atom-proton and atom-electron potentials, respectively,
and $r=(3v/4\pi)^{1/3}$.
In these evaluations, Eq. (\ref{e.gv_ideal}) was adopted for the species 
(electrons or protons) which let the nearest neighbor of an atom ($j,v$), 
while we approximate $g_{jv,c}(\omega) = \Theta(\omega - v)$ as the p.d.f. 
associated to the other perturber class.

From Eq. (\ref{e.Njv}) and the division of atoms into two groups (according 
to the type of nearest perturber), the internal partition function of the 
atomic hydrogen is given by
\begin{equation} \label{e.IPF}
Z = \sum_j g_j e^{-\beta \epsilon_{j} } \int_0^\infty
e^{-\beta \gamma v} \left( \frac{  e^{-\beta \Delta \epsilon_{jv}^{(e)} } +
e^{-\beta \Delta \epsilon_{jv}^{(p)} } } 2 \right) dv \,.
\end{equation}
This partition function includes a {\em state sum} over the physical space
as a direct consequence of the configurational parameter $v$ used in
the gas description. When the interactions are negligible 
($\Delta \epsilon_{jv}^{(e)}=\Delta \epsilon_{jv}^{(p)}=0$), the present 
partition function reduces to $Z = Z_0 /n$ where 
$Z_0 =\sum_j g_j e^{-\beta \epsilon_j}$
is the conventional form of the internal partition function of a perfect gas. 

\begin{figure}
\scalebox{0.45}{\includegraphics{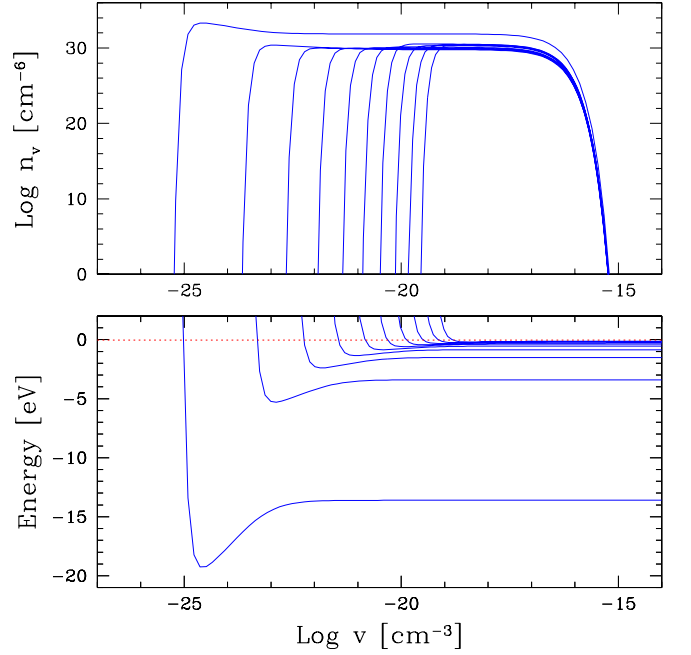}}
\caption{\label{f.feos_e} Double density ({\em above}) and energy 
({\em below}) curves of the first ten levels of hydrogen atoms (solid lines
from left to right), which have an electron as the nearest neirghbor. 
The results are shown as a function of 
the available volume per atom, and correspond to $T=15000$ K and mass 
density $\rho = 10^{-6}$ gr cm$^{-3}$. The dotted line shows the energy 
continuum edge for electrons and protons.}
\end{figure}
Populations and effective internal energies of atoms as functions of the 
available volume per particle are calculated for a gas with a
temperature $T=20000$ K and mass density $\rho=10^{-7}$ gr cm$^{-3}$. 
This is representative of conditions found in the atmospheres of lukewarm 
white dwarfs (effective temperature around $T_{eff}=15000$ K).
Non-ideal effects on equations of state are negligible and, particularly, 
$\beta \gamma = n = 10^{17.07}$ cm$^{-3}$. 
A standard evaluation of the chemical equilibrium yields $n_e=n_p=10^{16.76}$ 
cm$^{-3}$ and $n_H=10^{15.31}$ cm$^{-3}$.

\begin{figure}
\scalebox{0.45}{\includegraphics{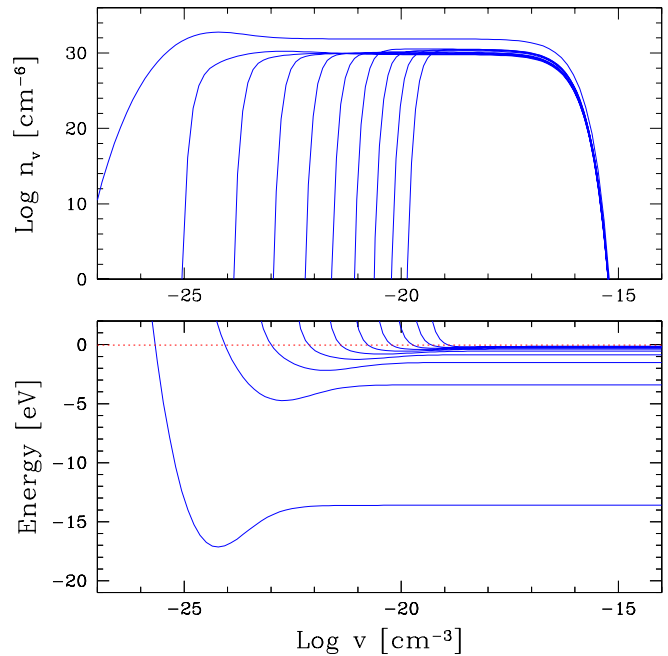}}
\caption{\label{f.feos_p} Same as Fig. \ref{f.feos_e}, but for atoms with a
proton as the nearest neighbor.}
\end{figure}

The effective internal energy of an atom is composed by the sum of its 
electronic energy $\epsilon_{j}$ (that chosen of an unperturbed atom) and 
the interaction energy $\Delta \epsilon_{jv}$ which condenses the plasma 
effects. 
Lower panels in Figs. \ref{f.feos_e} and \ref{f.feos_p} show the effective 
internal energy as a function of its available volume for the lowest atomic 
levels. Atoms with great volumes are, in practice, unperturbed and their 
internal energies are very close to those of an isolated atom.
As $v$ decreases the atoms become progressively more bound due to the
attractive long-range interaction with charges. This interaction has a 
stronger $v$-dependence for high-lying levels due to a greater shearing of the
electronic wavefunction, and accordingly, a greater polarizability. 

The Coulomb repulsions and Pauli's exclusion effects preclude atoms from having 
very low available volumes. These short-range interactions increase the
effective internal energy of atoms and bound states successively merge into 
the continuum with decreasing the available volume. 
Differences between energy curves of Figs. \ref{f.feos_e} and \ref{f.feos_p}
at low $v$ are due to the fact that Coulomb repulsion between protons 
[Eq. (\ref{e.rep_p})] is appreciably softer than the electron-atom 
interaction [Eq. (\ref{e.rep_e})]. In particular, atoms with an electron as 
the nearest neighbor have an abrupt increasing of interaction energy at low 
available volumes.
Notice that linear Stark effects are not taken into account because we use 
energy eigenvalue ($\epsilon_{j}$) of unperturbed particles.
However, since the linear Stark effects only spread over the levels 
without shifting their centers of gravity, the main effect on the internal
energy of the atoms is due to shifts caused by plasma polarization and 
short-range interactions considered in $\Delta \epsilon_{jv}$.

Density curves in Figs. \ref{f.feos_e} and \ref{f.feos_p} (upper panels)
show that atoms exist only for available volumes within a certain domain, 
which depends on the internal state $j$. 
The double densities take maximum values around the minimum of the 
internal energy curves. The absence of particles with very low volumes $v$
is due to the fact that atoms with high internal energies are not stable. 
Indeed, according to the energy curves, an atom with very low available 
volumes is so strongly perturbed that its electron becomes free in the plasma.
Atoms in high-lying bound states are more strongly affected by the presence 
of plasma electrons and ions, because their electrons are loosely bound.
Since the Coulomb repulsion $u_C$ is softer than the potential $u_{12}$,
atoms with a proton as the nearest neighbor may survive at lower available
volume than those with an electron as the nearest neighbor. However, this
difference does not have an influence on the total populations of each 
group of atoms. 

The population of atoms in all internal states declines at high $v$ volume
($v\approx 10^{-15.5}$ cm$^{-3}$) because actually there are no available
volumes per particle much greater than the mean volume per particle in the gas, 
$\langle v \rangle=n^{-1}=10^{-16.07}$. Mathematically, this is
a consequence of the term $\exp (-\beta \gamma v)$ in the double density 
expression [Eq. (\ref{e.Njv})]. Such factor plays the role of a
probability of occupation of available volumes. 
Large volumes $v$ represent single-particle, configurational states 
with smaller occupational probability than the little ones.

\begin{figure}
\scalebox{0.40}{\includegraphics{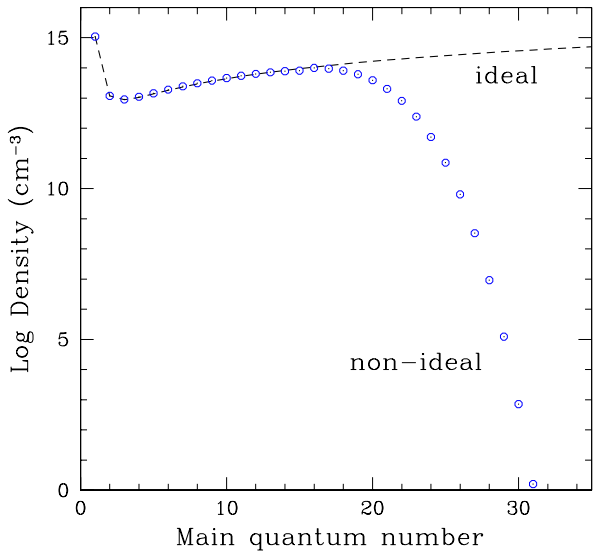}}
\caption{\label{f.feos_j} Populations of atomic levels as a function of the
quantum main number for $T=20000$ K and $\log \rho = -7$.
Open circles show results for atoms with an electron as the nearest neighbor,
while points correspond to atoms with a proton as the n.n. (actually both 
atom groups have very similar populations).
The dashed line shows results based on an ideal gas (i.e., for 
$\Delta \epsilon_{jv}^{(e)}=\Delta \epsilon_{jv}^{(p)} \equiv 0$).}
\end{figure}
Fig. \ref{f.feos_j} shows total densities (i.e., $v$-averaged) of atoms in 
differents internal states, together with predictions obtained for 
non-interacting particles. 
Low-energy bound states have essentianly unperturbed populations.
On the contrary, atoms in high energy levels are so strongly perturbed that 
their equilibrium populations decrease with respect to a hypothetical gas of 
unperturbed atoms. Under current conditions, states with main quantum  
numbers larger than $\approx 30$
are not populated because the atom, in order to survive, requires a 
volume $v$ much larger than the mean volume per particle and the probability 
of existence of such space becomes very small. 
Notice the smooth dependence of the non-ideal effects on atomic density with 
the quantum main number. This contrasts with evaluations of internal 
partition function based on an (often used) abrupt cutoff at certain upper 
level. The present evaluation of internal partition function contains useful
advantages derived from the free-energy minimization technique:
(i) convergence of the internal partition function, (ii) incorporation of 
particle-interaction models in a consistent statistical mechanical way.

A variety of gas models have been developed to compute the equilibrium state
of astrophysical fluids, and which include the influence of non-ideal gas 
effects upon the internal partition function of bound states 
\cite{Fonta, gasmodel, potekhin}. However,
to our knowledge, this is the first time that combined and self-consistent 
evaluations of atomic populations and internal effects on bound states are 
performed differentiating groups of atoms under different plasma-perturbations.
We consider this as an important advantage of the present formalism, which
can be useful for more elaborated models.
Notice also that the data derived from the population equation (\ref{e.Njv}) 
could be used to set up thermodynamic functions and to obtain optical 
quantities. This is not however within the scope of this paper.

\section{Conclusions} \label{s.conclus}

Classical statistics of gases in equilibrium has been extended taking
into account a new one-particle variable which assigns a space $v$ to each
particle. We call this variable the available volume per particle since
it roughly represents the free space around it.
We develop the theory in terms of a Helmholtz free-energy model for
one-component gases at low densities where the particle interactions are 
assumed to be binary.
The equilibrium state is calculated from the energy minimization with respect 
to the occupation number $N_v$ of $v$-states, subjected to particle number and
volume constrains. 
The volume-closure introduces a Lagrange's multiplier, here called $\gamma$,
which regulates the space partition (i.e., $N_v$) and is equivalent to the 
pressure minus the density of interaction energy in the gas. 
A universal expression [Eq. (\ref{e.Nv})] gives the occupation number $N_v$ 
for dilute gases at equilibrium. When the pair potential is specified, $N_v$ 
contains both thermodynamics and structure information of the fluid.
We have illustrated the application of the theory to the perfect gas and 
to fluids 
with infinitely repulsive interactions, for one, two and three dimensions.
All thermodynamic and geometric-structure properties such as the nearest 
neihgbor distribution function of these systems have been correctly reproduced. 
For hard bodies in two and three dimensions, the results are valid at low 
densities. It was also shown that the Van der Waals model can be derived from 
general K\^ac potentials in a straighforward way.

The identification of $v$ has been perfomed from the application of the 
theory to specific cases.
For noninteracting particles the available volume is equivalent to the 
spherical volume enclosed by the distance between the centers of the particle
and its nearest neighbor.  
In hard-body fluids, the core-repulsions reduce the value of $v$ with 
respect to that for fully penetrable particles. At the limit of low densities,
the explicit form of $v$ for hard bodies is derived from the virial theorem.
Hard particles provide a guidance for the treatment of soft-cores present in
realistic fluids. In these systems, we expect that the allowed values of $v$ 
will be a compromise between values associated with noninteracting particles 
and those corresponding to hard particles. Additional work should be 
done on this topic.
On the other hand, there is no evidence that the identification of $v$ can 
be modified by attractive interactions. This is reasonable because attractive 
forces do not restrict the accessible space to a particle.

For non-interacting particles $N_v$ follows an exponentially decreasing law, 
such that most particles (a fraction $1-e^{-1}\approx 0.63$) 
have spaces $v$ lower than the mean volume per particle ($1/n$).
In conexion with the structure of fluids, we have showed that the double 
density $n_v$ could be a useful tool to characterize the order in fluids, 
particularly, through the standard deviation of the available volume $v$.

The main application of the formalism concerns with the atomic component 
of a partially ionized hydrogen gas.  
We found a statistically coherent and detailed evaluation
of populations and internal energy changes of atoms as a function 
of the available volume per particle.
Preliminary calculations show that the distinction of atoms under
different plasma-perturbations emerges naturally from the theory.
This finding provides sufficient confidence for further extension of the 
formalism to consider a complete description of gaseous mixtures, 
where non-ideal effects on optical properties may be investigated.

\begin{acknowledgments}
Thanks are due to J. Cant\'o, L. Garc\'{\i}a-Col\'{\i}n, D. Page, 
S. Sahal-Br\'echot and J. Zorec for their help, comments and discussions on 
the research reported in this paper. 
I also want to thank to L. Garc\'{\i}a-Col\'{\i}n for a careful 
reading of the manuscript and valuable comments,
J. Zorec for their kind hospitality during me stay in the Institut 
D'Astrophysique of Paris, and Jana Benda for improving the language of the 
manuscript.
\end{acknowledgments}


\begin{thebibliography}{99}
\bibitem{foot1} Due to the wide field of interest, there is an extensive 
terminology. Dirichlet regions, Wigner-Seitz cells, Thiessen figures,
Vorono\"{\i}-Delaunay and Bernal's polyhedra are some of the terms usually  
employed in the literature to refer to regions resulting of a variety of space 
partitions.
\end{thebibliography}

\end{document}